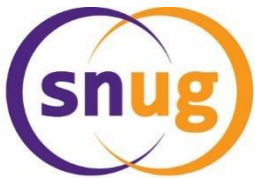

# Constraining the Synopsys Pin Access Checker Utility for Improved Standard Cells Library Verification Flow

Yongfu Li, Chin Hui Lee, Wan Chia Ang,
Kok Peng Chua, Yoong Seang Jonathan Ong, Chiu Wing Colin Hui

GLOBALFOUNDRIES

## ABSTRACT

While standard cell layouts are drawn with minimum design rules for maximum benefit of design area shrinkage, the complicated design rules begin to cause difficulties with signal routes accessing the pins in standard cell layouts. Multiple design iterations are required to resolve routing issues, thus increasing the runtime and the overall chip area. To optimize the chip performance, power and area (PPA) and improve the routability, it is necessary to consider the pin accessibility during standard cell development phase so that each cell is designed to maximize the number of feasible pin-access solutions available to the router. As part of the Synopsys IC Compiler Library Preparation Reference Methodology, the Synopsys Pin Access Checker (PAC) reports DRC violations associated with the standard cell. Based on Synopsys PAC's methodology, we demonstrate several methods to improve the probability of detecting pin accessibility issues, such as reducing the number of cells required for each Synopsys 'testcell', increasing the complexity of the pin connectivity assignment and recommending the router constraints.

# Table of Contents

# Table of Figures

## Table of Tables

# I. Introduction

Every successful scaling of the CMOS technology comes along with an exponential increase in the number of design rules. The design rule check (DRC) can no longer guarantee 100% pattern printability in the nanometer CMO technology and design-for-manufacturability (DFM) compliance checking is required to identify manufacturing weak-points and prevent catastrophic errors such as open (necking) and shorts (bridging) issues [1]. With the additional of mandatory physical verifications in every new technology generation, the development resource for the intellectual property (IP) libraries has become astronomical costly. Therefore, early assessment of the design restrictions imposed on the circuit design is absolutely required.

In every technology, standard cell and input-output (I/O) IP libraries are the first and foremost design foundation libraries to construct the place-and-route (P&R) digital circuit design. The quality of the standard cell layouts has a direct impact on the chip design's area and manufacturability. Today, one of the main challenges limiting the quality of the standard cell library is the placement of the pin locations.

As the standard cell layouts are drawn with minimum design rules to achieve minimum design rules for the maximum benefit of design area shrinkage, the number of routing tracks has begun to decrease from 12-track in 65-nm to 7.5-track in 14-nm. As a result, it was reported that the modern router tools have difficulties with signal routes accessing to the standard cell pins in the design [2]. In particular, Hsu, et al. reported that the difficulties with pin-access have severely degraded routing resource estimation accuracy with the convention global and detail routing model [3]. Multiple design iterations are required to resolve the routing congestion and pin accessibility issues, or even required in a change in the design core utilization, thus resulting in an increase in the overall chip area [4]. The experiment has shown that the smallest chip area is not achieved by standard cells with the smallest area [2].

To achieve optimal design and to mitigate routing congestion, it is important to consider pin accessibility during the layout development phase. Synopsys has introduced the Synopsys Pin Access Checker (PAC) utility in the library preparation reference methodology, which aims to identify problematic pin location in the standard cell layouts [5]. This motivates us to explore this new methodology and to improve the detection coverage with our proposed design constraints with the utility tool.

The rest of this paper is organized as follows. Section II explains the present challenge in identifying the pin accessibility issue. Section III presents the overview of the Synopsys PAC flow. Section IV presents several methods to increase the detection coverage for problematic pins in the standard cell libraries. Finally, Section V presents the result using our proposed constraints and conclusions are given in Section VI.

## II. Present methodology to identify pin accessibility issue.

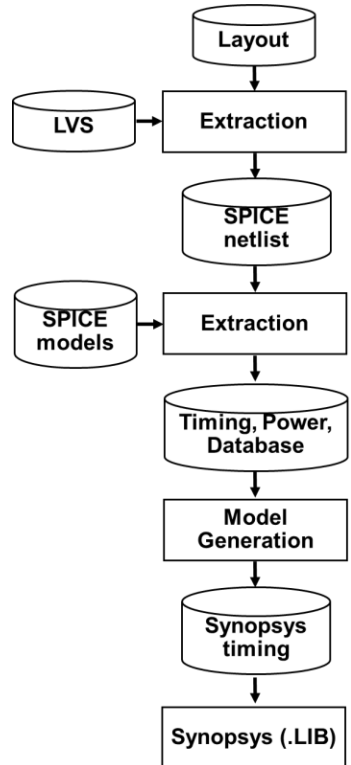

Figure 1. Timing Library Generation Flow.

Based on the conventional methodology, the library development team has to perform physical verification checks on a P&R digital circuit design in order to identify and fix the standard cells with pin-access difficulties. This involves the characterization of the standard cells' layout, followed by the generation of the Synopsys timing liberty library format (.LIB), as illustrated in Figure 1. The Synopsys reference library milkyway database is created with references from the the Synopsys IC Compiler (ICC) Library Preparation Reference Methodology [5]. Both inputs, the timing library and milkyway database, are used to translate the Verilog files into the digital circuit design through the synthesis and P&R design methodology.

The presented physical verification and layout correction methodology involves various groups of engineers such digital circuit designers, P&R layout engineers and library development engineers, to perform all the required tasks. Additional resources and time are required to perform multiple iterations for any changes in the layout or any new cells added to the library. Another disadvantage is that the synthesized netlist might not achieve hundred percentage standard cells' coverage. Thus, the solution is less likely to be adopted for incremental development work. There is a need to simplify the present solution and provides an easy-to-use utility for library physical layout designer to perform all the tasks with hundred percent verification coverage. The simple-to-use approach reduces the barriers to adopting new tool, especially the layout engineers who may not be well-versed with programming and digital design flow.

## III.    Synopsys Pin Access Checker (PAC) Flow

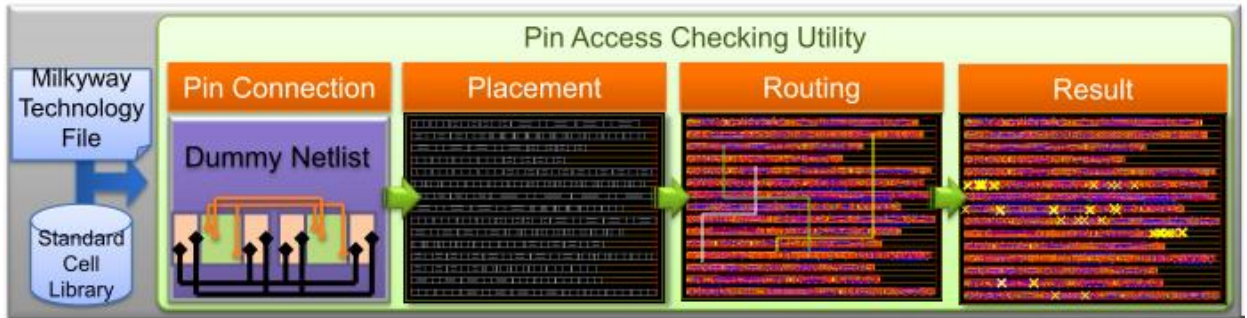

Figure 2. Synopsys PAC utility workflow.

To address the pin accessibility challenge with a simplified solution, Synopsys has provided the Synopsys PAC utility as part of the Synopsys IC Compiler (ICC) library preparation reference methodology [5] and the concept of the Synopsys PAC utility is summarized in Figure 2. The utility enumerates all combinations of the standard cells and generates a Verilog gate-level netlist and a design exchange format (DEF) file which are essential to define the logical connections and physical placements of the generated design, respectively. Synopsys termed this circuit-like design as 'testcell' method. The utility also leverages on the Synopsys ICC zroute capability to perform automated routing and design rule check (DRC) verification. The DRC violations help to identify the pin locations associated with routing congestion issues. Therefore, this utility simplifies the physical verification flow for pin accessibility check without the need for timing library and timing constraints.

The Synopsys PAC utility provides four different modes of checking pin accessibility, namely (1) single_cell_only, (2) cell_by_cell_only, (3) all_combo_in_one_cell_only and (4) all [5], [6]. The first two modes allow library development engineers to perform incremental verification during the standard cell layout development phase and the latter two modes can be performed on the final library before silicon validation phase. The latter two modes create the top cell with the entire individual 'testcells' generated through the modes (1) and (2). The multiple-tier verification approach helps to perform verification with limited runtime overhead at the various phases of library development. Thus, the library development engineers will be able to correct the problematic cells before silicon validation phase.

In [6], Aupoix, et al. evaluated the Synopsys PAC utility with its default constraints on several standard cell libraries, which only enables them to identify one problematic pin issue among the libraries. Furthermore, the initial pin accessibility problem with the testcell was later resolved with rerouting. This is because since the detail router is based on the heuristic algorithm, the final routed design highly depends on the placement of the standard cells and the number of routing iterations. Therefore, the number of problematic pins identified through this methodology depends on the final routed design.

Based on the previous learnings in [6], we attempt to explore several methods to constraints the placement of 'testcell' and it's routing, thus exacerbating the pin accessibility issue. With the proposed methods, we aim to improve the probability of detecting pin accessibility issues.

## IV.    Improvability methods in the Synopsys PAC Utility

In this section, we discuss in more details methods that we can explore to improve the Synopsys PAC utility. We will focus on three key areas of improvements, namely our proposed 'testcell' architecture and our proposed pin connectivity assignment and the routing constraints.

### A.  Design Implementation

Synopsys releases the PAC utility in the form of compiled binary format to the customers. As such, it is not possible to make a direct modification to the utility in order to demonstrate our improveability methods. Based on the concept of the Synopsys PAC utility as discussed in Section III, we have implemented a simplified prototype version of the PAC utility, to demonstrate our improvability methods. Our "PAC" prototype utility is based on a customized Tcl procedure script, written for the Synopsys ICC tool.  The associated ICC commands needed to implement the flow are summarized in Figure 3 and the customized Tcl procedure can be broken down into the following phases.

- A.   Standard cell profiling
- B.   Verilog and DEF files generation
- C.   Cell Placement and routing
- D.   Design Rule Verification.

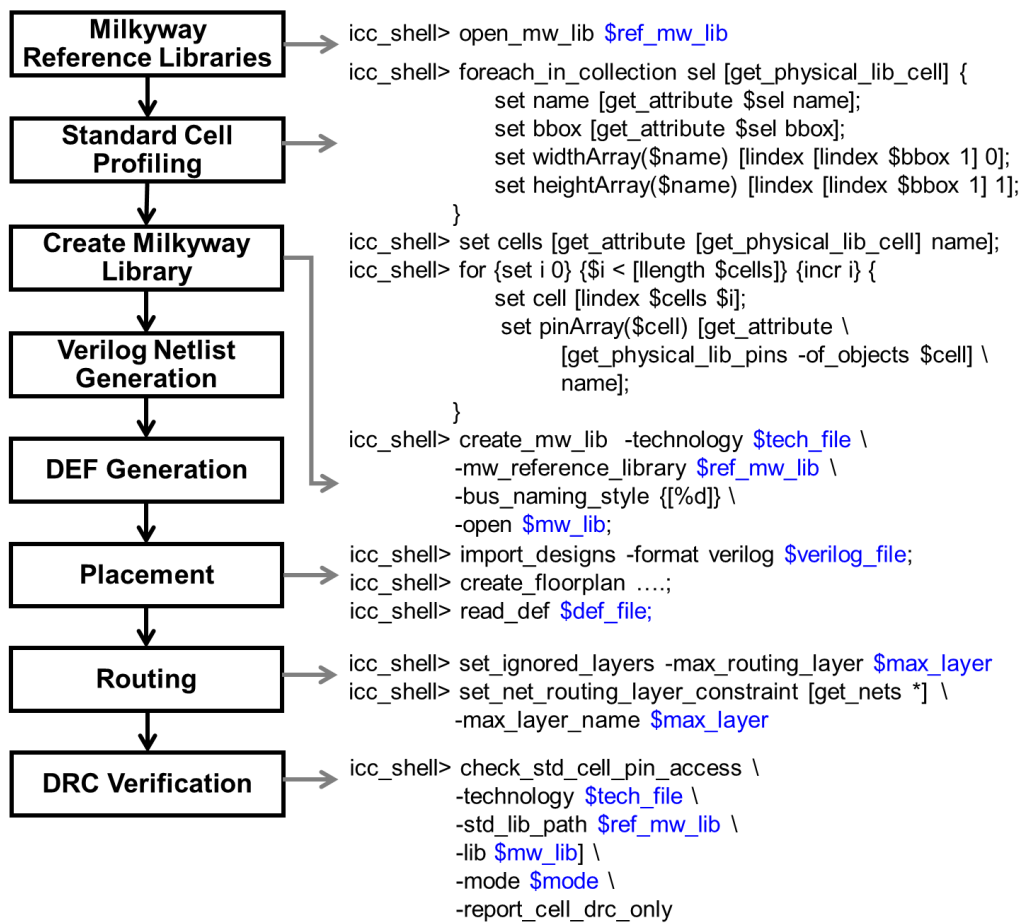

Figure 3. Workflow of the "PAC" prototype utility with the associated ICC commands.

*Constraining the Synopsys Pin Access Checking Utility for*
*Improved Standard Cells Library Verification Flow*

## B. Our proposed 'testcell' architecture

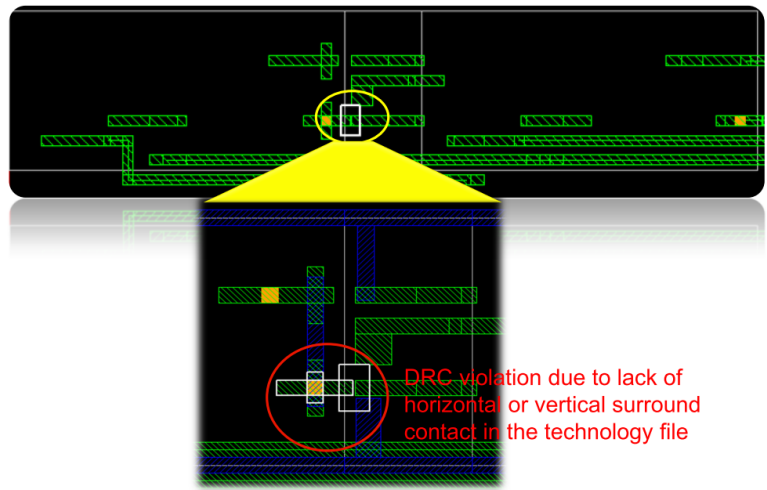

Figure 4. Metal 2 Spacing violation due to Metal 2 – Via 1 inserted at the pin location.

Besides the wire routing congestion problem, pin-access issue can also results in DRC violations at the cell boundaries. An example of such scenario is shown in Figure 4 where metal 2 spacing violation has occurred at the standard cell boundary due to the insertion of the metal 2 and via 1 at the pin location. To achieve hundred percentage cell boundaries coverage, the layout test case needs to include all the possible cell orientations with different cell abutment conditions. In the chip layout design, there are four possible orientations of a standard cell that can be placed in the layout, as shown in Figure 5. It is unlikely that the test case generated from the synthesis flow is able to achieve such coverage.

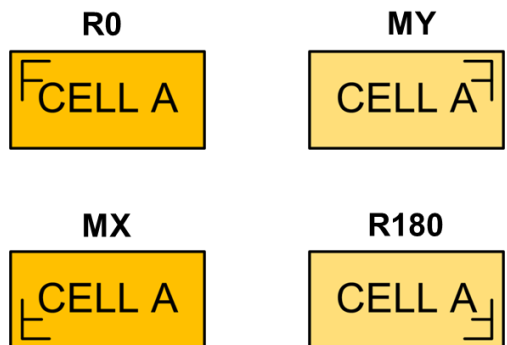

Figure 5. Four possible legal orientation of standard cell placement in the chip design.

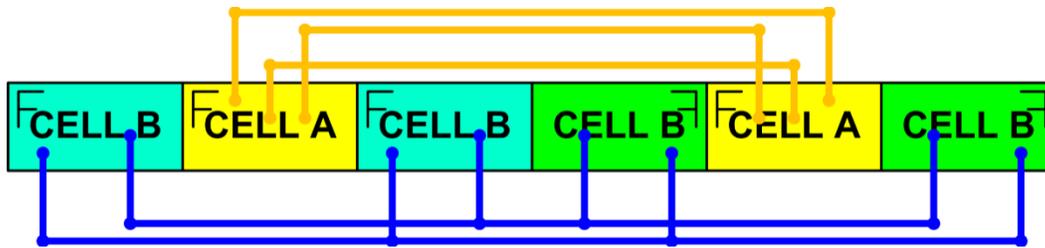

Figure 6. The schematic illustrates the simplified version of the Synopsys 'testcell' method.

Synopsys has introduced the 'testcell' method, as illustrated in Figure 6, where the target cell A (with R0 orientation) is placed in the middle with two cell B (with R0 orientations) are abutted on its right and left sides and the second set is similar to the first set, but the two cell B are flipped in the horizontal direction (with MY orientations). Thus, the placement in the Synopsys 'testcell' has considered all the possible combinations using the minimum number of cell required.

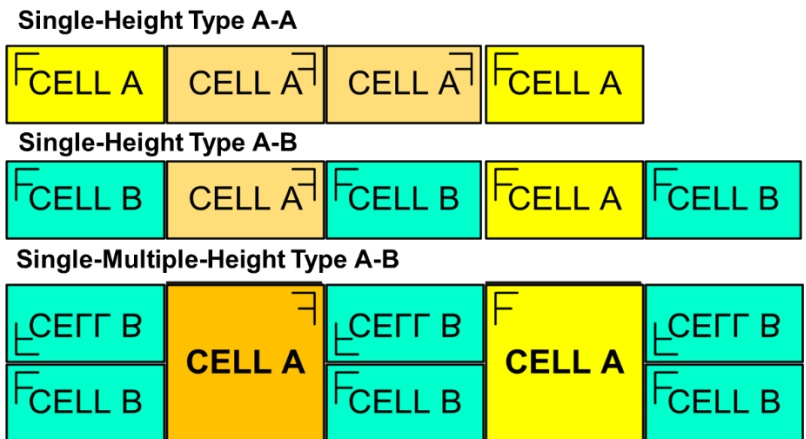

Figure 7. Proposed area-efficient cell abutment placements.

Although the Synopsys 'testcell' provides 1.3 times reduction in the number of cells compared to the conventional method, there is a need to further optimize the testcell in order to reduce the database and runtime significantly. Without the loss of generality, we proposed three types of area-efficient cell abutment placements, which reduce the number of required standard cells compared to the Synopsys 'testcell'. In order to achieve a significant cell reduction, we have to consider the following scenarios:

- Single-height type A-A – Single-height side-to-side standard cell abutment with identical cells.
- Single-height type A-B – Single-height side-to-side standard cell abutment with different cells.
- Single-multiple-height type A-B – Single-height side-to-side standard cell abutment with the multiple-height cell.

The breakdown of different scenarios allows us to further optimize the number of cells required for each abutment condition. Although this work did not consider multiple-height-multiple-height Type A-B, the methodologies can be extended from the single-multiple-height Type A-B abutment placement. A detail explanation of our implementation is discussed in [7].

To illustrate the benefits of our proposed method, let us consider a standard cell library with N number of single-height cells. Our method reduces the total number of cells from 8*$N$ + 8*($N$-1)*$N$ to 4*$N$ + 2.5*($N$-1)*$N$, whereas the Synopsys 'testcell' method only reduces to 6*$N$ + 6*($N$-1)*$N$ *cells*. In one of our 14-nm technology IP library, the number of single-height standard cells is approximately 1000 cells. The conventional method requires a total of 8,000,000 cells to achieve hundred-percent coverage. The total number of cells used in the Synopsys 'testcell' method and our method is 6,000,000 and 2,505,500 cells, respectively. Therefore, our method provides close to 2.4 times reduction compared to the Synopsys 'testcell' method. Table 1 summarizes the total number of cells for each method.

Table 1. The number of cells required for each method.

| Number of Cells | Conventional Method | Synopsys 'testcell' | Our method |
|---|---|---|---|
| N | 8*$N$ + 8*($N$-1)*$N$ | 6*$N$ + 6*($N$-1)*$N$ | 4*$N$ + 2.5*($N$-1)*$N$ |
| E.g 1000 | 8,000,000 | 6,000,000 | 2,505,500 |
| Reduction | 1x | 1.33x | 3.2x |

## C. Pin connectivity assignment

The pin connectivity assignment in the Synopys 'testcell' is illustrated in Figure 6. The pins of the cell A and B are connected to the corresponding pins in the cell A and B, respectively. Since the cells in the Synopsys 'testcell' are placed in a single row with either R0 or MY orientations, the pins are aligned in the same routing tracking. Due to its simplicity in the Synopsys 'testcell' pin connection, the physical signal routes in all the four modes are very similar. This is evidently shown in Figure 8, the layers ME2 and ME3 routed nets in (a) single_cell_only mode are almost identical to (b) all_combo_in_one_cell mode, except for the net highlighted in 'red' box. In our opinion, Synopsys PAC utility verification modes (1) and (2) would be sufficient for the pin-accessibility assessment.

Figure 9 details the layout view of the Synopsys 'testcell' and the routed signal layers. We have observed that each signal only requires two to three via transitions to connect the pins together. Since the horizontal distance between the two pins are within two to three cell's width apart and the pins are aligned in the same routing track, this will be a trivial solution for the router tool. To increase the routing congestion and to exacerbate the pin accessibility issue, we propose to randomly connect the pins in our 'testcell'. This increases the routing distance and the signal routes are crisscrossed all over the 'testcell'. The routed 'testcell' design will be close resemblance to routings in the real design.

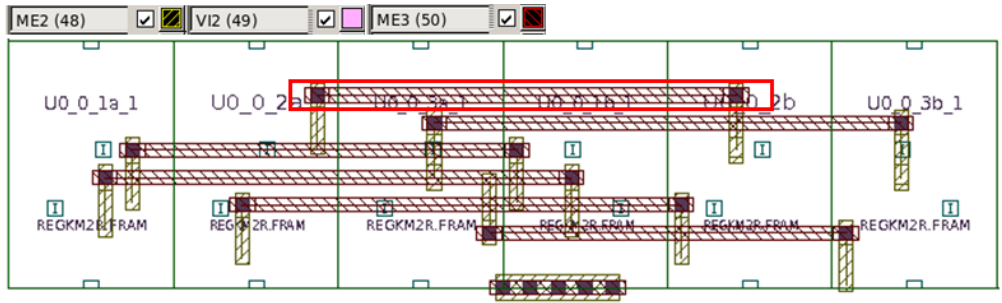

(a) ME2&ME3 routed nets with VE2 vias locations in single_cell_only mode

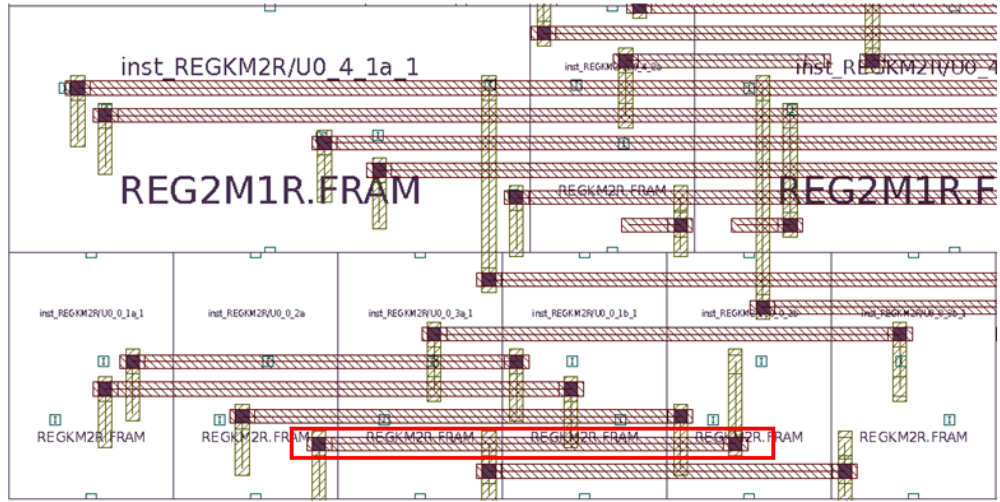

(b) ME2&ME3 routed nets with VE2 vias locations in all_combo_in_one_cell_only

Figure 8. An example of the Synopsys 'testcell' in the Synopsys layout environment (a) single_cell_only mode, (b) all_combo_in_one_cell.

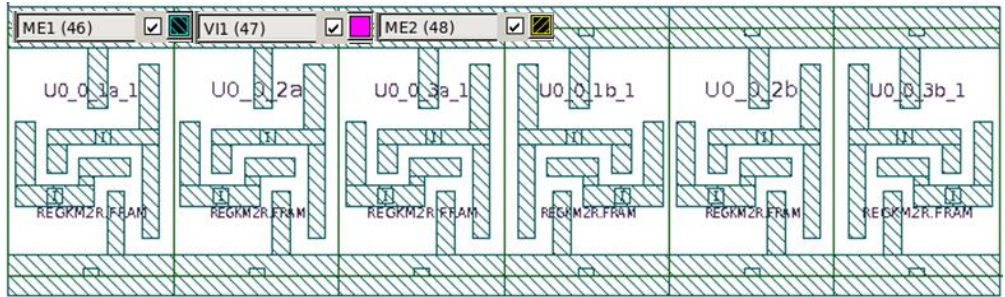

(a) Synopsys 'testcell' Placement

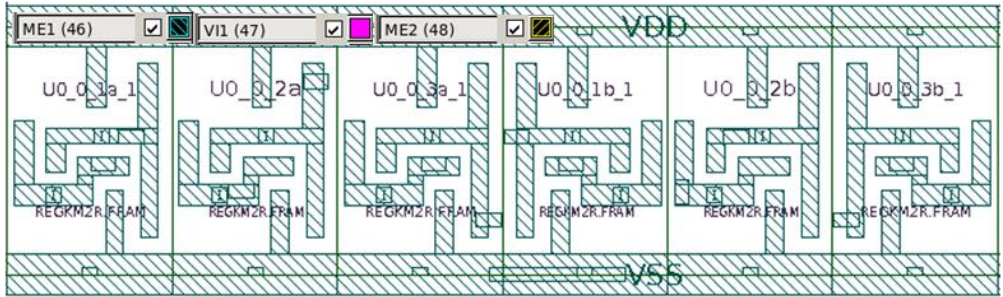

(b) ME1 routed nets

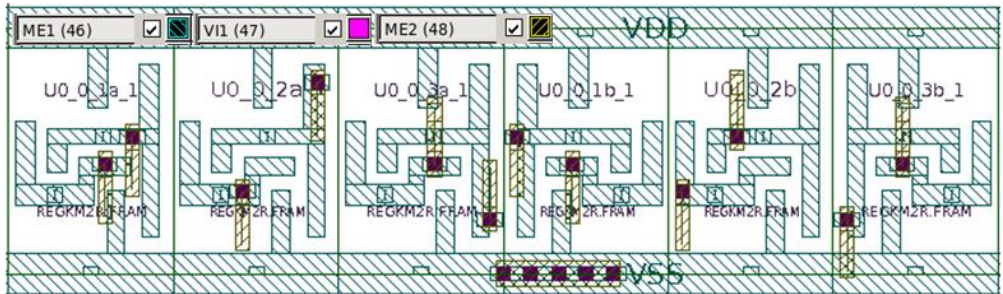

(c) ME1&ME2 routed nets with VE1 vias locations

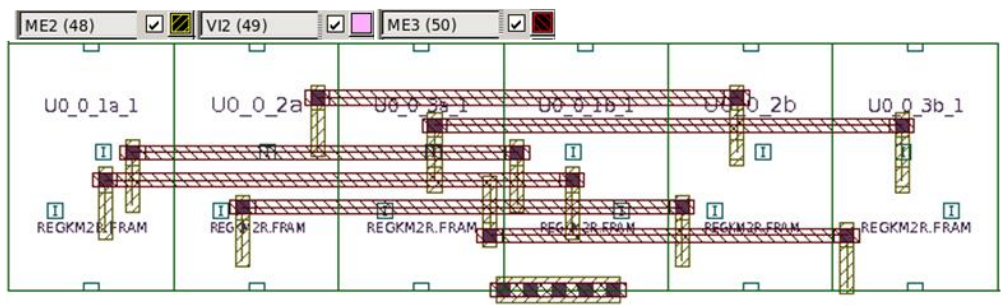

(d) ME2&ME3 routed nets with VE2 vias locations

Figure 9. An example of the Synopsys 'testcell' in the Synopsys layout environment (a) cell view of the Synopsys 'testcell', (b) ME1 layer signal routes, (c) ME2 and Vi1 layers are generated on the pins during the routing phase. (d) ME3 and Vi2 layers are generated to connect the pins together.

## D. Verilog and DEF files generation

To implement our proposed 'testcell' in the Synopsys IC Compiler, we have implemented the automated netlist and DEF file generation using our proposed utility. The Verilog netlist describes the logic cell and the pin connectivity as shown in Figure 10. For example, for the single-height type A-A placement, it requires four identical cell, which is described under the module scell_<typeA> Verilog netlist. The DEF file describes the physical layout of the 'testcell'. It requires the basic standard cell information such as width and height, which is stored in the Tcl arrays during the standard cell profiling phase. The orientation of each cell is derived based on the type of abutment placement. For example, as shown in Figure 11, for the single-height type A-A placement, the cell orientation is N – FN – FN – N.

**Single-Height Type A-A**
**Multiple-Height Type A-A**
module scell_<typeA> ();
    <typeA> U1 (…);
    <typeA> U2 (…);
    <typeA> U3 (…);
    <typeA> U4 (…);
endmodule

**Single-Height Type A-B**
module scell_<typeA>_<typeB> ();
    <typeB> U1 (…);
    <typeA> U2 (…);
    <typeB> U3 (…);
    <typeA> U4 (…);
    <typeB> U5 (…);
endmodule

**Single-Multiple-Height Type A-B**
module mcell_<typeA>_<typeB> ();
    <typeB> U1 (…);
    <typeB> U2 (…);
    <typeA> U3 (…);
    <typeB> U4 (…);
    <typeB> U5 (…);
    <typeA> U6 (…);
    <typeB> U7 (…);
    <typeB> U8 (…);
endmodule

Figure 10. Verilog netlist for different standard cell abutment.

```
Example of the generated DEF file

VERSION 5.6 ;
DESIGN TOP;
TECHNOLOGY ROUTE ;
UNITS DISTANCE MICRONS 1000 ;
COMPONENTS 4;
 - sinst_<typeA>\/U1 <TYPEA> + PLACED ( 0 0 ) N ;
 - sinst_<typeA>\/U2 <TYPEA> + PLACED ( 200 0 ) FN ;
 - sinst_<typeA>\/U3 <TYPEA> + PLACED ( 400 0 ) FN ;
 - sinst_<typeA>\/U4 <TYPEA> + PLACED ( 400 0 ) N ;
END COMPONENTS
DIEAREA ( 0 0 ) ( 600 0 ) ;
END DESIGN
```

Figure 11. DEF file to define the placement of the standard cell abutment.

The following ICC commands detail the floorplan implementation of the abutment placement:

```
create_mw_lib \
    -technology $tech_file \
    -mw_reference_libraray $mw_libs \
    -bus_naming_style {[%d]} \
    -open $lib; # Create the milkyway library
import_design –format Verilog $verilog_file; # Read in the Verilog file
create_floorplan ...; # Create the floorplan (Optional)
read_def $def_file; # Read in the DEF file
```

## E.  Routing Constraint

The technology file details the physical design rule constraints for the metal routing lines. To increase the routing congestion and to exacerbate the pin accessibility issue, Synopsys PAC utility provides the option 'route_option_file' that specifies the route options to be honored during the routing phase [5]. However, there is no clear guideline for the list of routing constraints needed.

In this work, we have experimented with various ICC routing constraints commands to create random blockage and routing nets to increase the routing congestion. This is list of commands can be used in the Synopsys PAC utility. The following ICC commands detail the routing constraints:

Step 1: Derive the routing layers information.
```
set layers [get_attribute [get_layers -filter is_routing_layer==true] full_name];
set horizontal_layer [get_attribute [get_layers -filter preferred_direction==horizontal] full_name];
set vertical_layer [get_attribute [get_layers -filter preferred_direction==vertical] full_name];
set via_layers [get_attribute [get_layers –filter is_routing_layer==true –filter layer_type==via] \
full_name];
foreach layer $layers { if {[lsearch $via_layers $layer] == -1} { append route_layers "$layer "; }; }
```

Step 2: Connects the power rails together.
```
derive_pg_connection -power_net VDD -power_pin VDD -ground_net VSS -ground_pin VSS;
preroute_standard_cells -fill_empty_rows \
    -advanced_via_rules \
    -nets "VDD VSS" \
    -route_type {P/G Std. Cell Pin Conn} \
    -extend_to_boundaries_and_generate_pins;
```

Step 3: Constraints the router to 2 routing layers.

```
set min_layer [lindex $layers 2]; # minimum routing layer
set max_layer [lindex $layers 4]; # maximum routing layer
set_preroute_drc_strategy -min_layer $min_layer -max_layer $max_layer;
set_ignored_layers -min_routing_layer $min_layer;
set_ignored_layers -max_routing_layer $max_layer;
set_net_routing_layer_constraint [get_nets *]  -min_layer_name $min_layer -max_layer_name $max_layer;
set constraints "";
for {set i 0} {$i <= [llength $route_layers]} {incr i } {
    if $i {
        append constraints "{[lindex $layers $i] false} "
    } else {
        append constraints "{[lindex $layers $i] true} "
    };
};
set_route_zrt_common_options  -freeze_layer_by_layer_name $constraints;
```

Step 4: Create random power straps (acts as random routing blockage).

```
set vertical_layer [get_attribute [get_layers -filter preferred_direction==vertical] full_name];
set ymax [lindex [lindex [get_attribute [get_die_area] bbox] 1] 1]; # size of the 'testcell'
set xmax [lindex [lindex [get_attribute [get_die_area] bbox] 1] 0]; # size of the 'testcell'

foreach layer $route_layers {
    if {[lsearch $horizontal_layer $layer] >= 0} {
        create_power_straps -direction horizontal \
            -start_at 0 -stop $ymax \
            -layer $layer -nets "VDD VSS" \
            -width [expr int(rand()*100)/500.000] \
            -step  [expr int(rand()*100)/50.000] \
            -configure step_and_stop \
            -extend_low_ends force_to_boundary_and_generate_pins \
            -extend_high_ends force_to_boundary_and_generate_pins \
            -keep_floating_wire_pieces
    } else {
        create_power_straps -direction vertical \
            -start_at 0 -stop $xmax \
            -layer $layer -nets "VDD VSS" \
            -width [expr int(rand()*100)/500.000] \
            -step  [expr int(rand()*100)/50.000] \
            -configure step_and_stop \
            -extend_low_ends force_to_boundary_and_generate_pins \
            -extend_high_ends force_to_boundary_and_generate_pins \
            -keep_floating_wire_pieces
    }
}
```

Step 4: Constraints the router to fix open and short errors only.

```
set_route_zrt_detail_options -report_ignore_drc [list "Diff net spacing" "End of line spacing" \
    "Diff net var rule spacing" "Same net spacing" ... ]; # List of drc constraints to be ignored
    during reporting
```

To specify the router to fix the open and short locations, we use the following ICC commands:

```
route_zrt_detail -incremental true -initial_drc_from_input true
```

In this work, we have used the option 'report_cell_drc_only' in the Synopsys PAC utility to summarize the DRC error types. A simple experiment is conducted on one problematic standard cell. As shown in Figure 12, the Synopsys PAC's DRC report only indicates that there is one DRC error in our proposed 'testcell'. This demonstrates that our proposed methods have improve the odds of detecting the pin accessibility issues in the standard cell. As shown in Figure 13, in our proposed 'testcell', cell_REGKM2R_yf, there is a ME1 spacing violation at the standard cell boundary due to the insertion of the metal 2 and via 1 at the pin location. The ME1 spacing violation is highlighted with the 'red' box.

It is noted that the accuracy of the DRC report is dependent on the layout content in the milkyway database's FRAM view. For example, if the FRAM view only contains the pin information without M1 metal layer, the extra M1 metal layers added to fulfil the enclosure condition might violate the M1 metal layer spacing violation at the cell boundary. Therefore, it is highly encourage for engineer to use the complete layout in the FRAM view for the verification.

```
=========================================================================
SCRIPT-Info: Printing DRC Summary ....
=========================================================================
########## 4 cells without DRC errors ##########
---------------------------    ---------  ------------------------------------------------     --------------------
Cell                           DRC count  Master Cells with DRC                                DRC Types
---------------------------    ---------  ------------------------------------------------     --------------------
auto__icGRt                    0
auto__icTrkAsgn                0
auto__icDRt                    0
cell_REGKM2R                   0
########## 1 cells with DRC errors ##########
cell_REGKM2R_yf                1                                                               {Same net via-cut spacing}
```

Figure 12. Synopsys PAC utility's DRC report, reporting DRC errors in the proposed 'testcell' compared to the Synopsys 'testcell'.

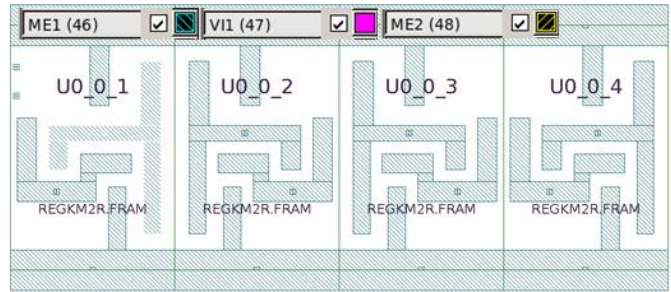

(a) Proposed 'testcell' Placement

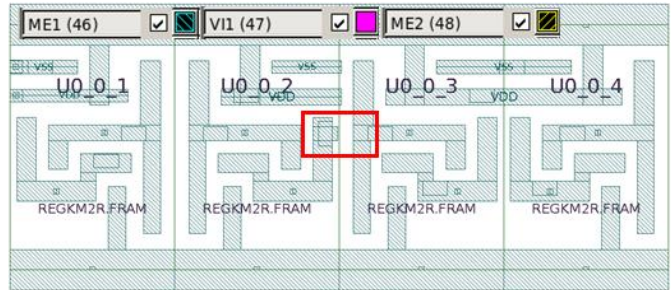

(b) ME1 routed nets

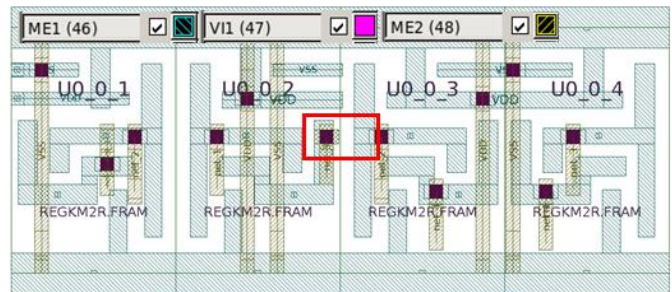

(c) ME1&ME2 routed nets with VE1 vias locations

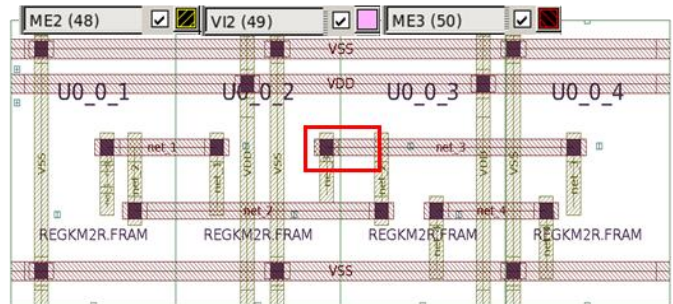

(d) ME2&ME3 routed nets with VE2 vias locations

Figure 13. An example of our 'testcell' in the Synopsys layout environment. The ME1 spacing violation is highlighted with the 'red' box. (a) Cell view of our 'testcell', (b) ME1 layer signal routes, (c) ME2 and Vi1 layers are generated on the pins during the routing phase. (d) ME3 and Vi2 layers are generated to connect the pins together.

# V.    Result & Discussion

The experiment was carried out on six sets of 14-nm standard cell IP libraries. The library information is listed in the Table 2. The first four set of libraries belong to 9-track height design, which is optimized for mainstream design while the remaining two set of libraries belong to 10.5-track height design, which is optimized for speed critical design. 'library1' and 'library6' belongs to the standard threshold standard cell IP library. 'library2', 'library3' and 'library4' belongs to different engineering versions of the mixed-threshold standard cell IP library. As shown in Figure 14, 'library1' contains about eighty percent of the standard cells fall within ten times of the minimum standard cell width. As shown in Figure 15, the mixed-threshold standard cell IP library, 'library2' contain about thirty percent of multiple-height standard cells.

In our experiment, we have implemented our methods and the recommended methodology presented in [6] using the ICC Tcl language. The metric used to compare the two utilities are the run-time, the size of milkway database and the number of problematic standard cells. The run-time refers to the total compute time taken for the verification to complete on our Linux workstation with Intel 2.7-GHz 8 Core Duo CPU and 128-GB of memory while restricting the router to a four thread for consistency in comparison. We constraints the routing to two metal layers (M2 and M3) and the routing directions of M2 and M3 are horizontal and vertical, respectively.

The results are shown in Table 3. As shown in this table, our utility incurred additional run-time and increased milkway databases size compared to the recommended methodology. The main contributing factors to the increased in the run-time and database are explained as follow:

A.  Our algorithm is implemented in Tcl language and therefore, the implemented algorithms are executed with a single-thread process compared to the Synopsys PAC utility.

B.  Our proposed random pin connectivity assignment increases the Verilog file generation compared to the Synopsys PAC utility.

C.  In this work, we have saved all different version of the milkyway information separately. Therefore, it is expected that there will be a significant increase in the database. This is to allow us to understand the DRC errors changes with different re-routing iterations.

D.  Since our method will be able to identify more problematic cells, it is expected that the increased database is partly due to the storage of DRC errors information.

To improve on the run-time, our algorithm can be implemented in the C/C++ language so that we can leverage on the computer parallel threading resource to generate the Verilog and DEF files simultaneously as the individual files are independent from each other. The additional runtime from the routing phase is expected due to the tighter routing constraints compared to the default constraints. To optimize on the size of database, we will discard the incremental database versions.

It should be noted the results show that our proposed utility is able to identify more problematic standard cells. As expected, there will be more problematic cells in the 9-track libraries compared to the 10.5-track due to the decreased opportunity in the routing space and pin placement in the former libraries. In general, we have noticed that most of the DRC violations can be categorized under spacing violations. Examples of the DRC violations occurred at the pin locations are illustrated in Figure 16 (Local double pattern cycle violation), Figure 17 (Self aligned via cut spacing) and Figure 18 (Diff net spacing).

Table 2. Summary of the Standard cell libraries used for evaluation.

| Library Name | No. of Tracks | No. of cells. | No. of single-height cell | No. of multiple-height cell |
|---|---|---|---|---|
| library1 | 9 | 108 | 108 | 0 |
| library2 | 9 | 85 | 64 | 21 |
| library3 | 9 | 93 | 64 | 29 |
| library4 | 9 | 97 | 67 | 30 |
| library6 | 10.5 | 108 | 108 | 0 |

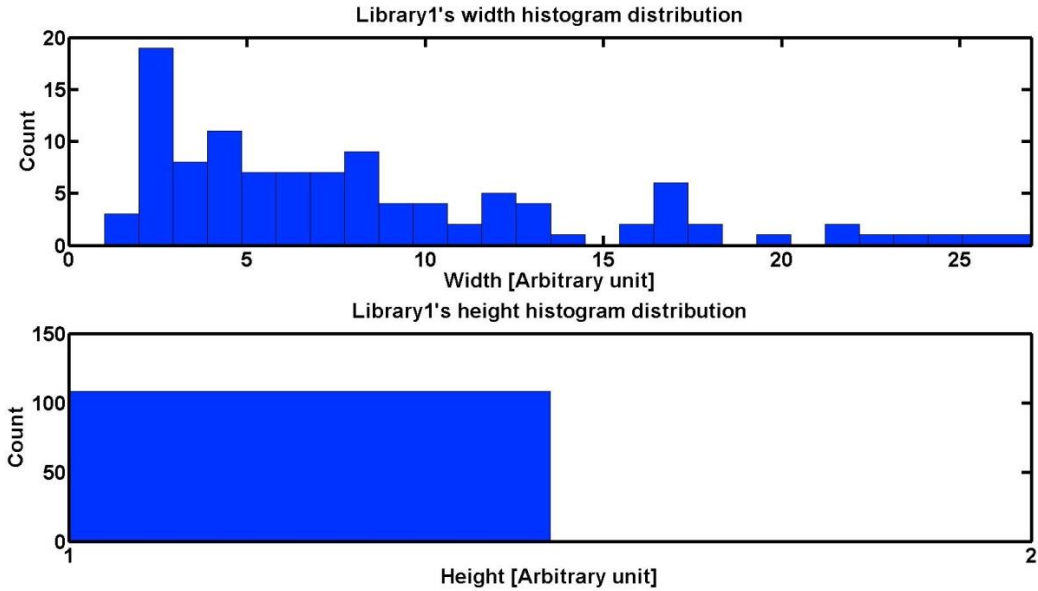

Figure 14. Normalized histogram distribution for the 9-track standard cell library1.

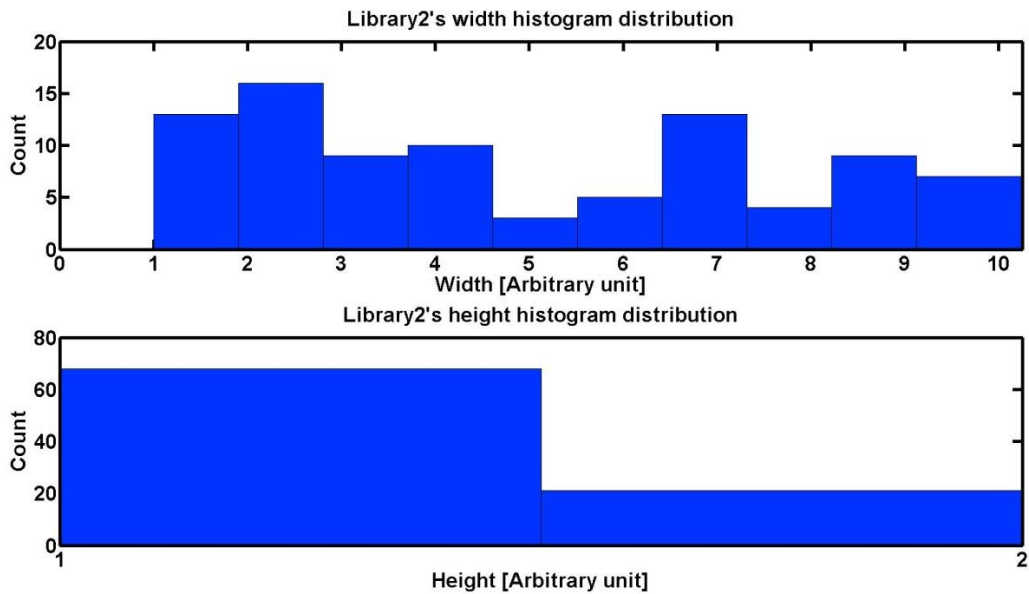

Figure 15. Normalized histogram distribution for the 9-track mixed-threshold standard cell library2.

Table 3. Comparison result between our proposed method and present recommended method.

| Library Name | Recommended methodology presented in [6] | | | Our proposed method | | | |
|---|---|---|---|---|---|---|---|
| | No. of cells with remaining DRC Route issues | Run-time [hours] | Database [Mb] | No. of cells with remaining DRC Route issues | Run-time [hours] | Database with incremental result [Mb] | Database without incremental result [Mb] |
| **library1** | 3 | 5.6 | 218 | 14 | 106 | 2,700 | 155 |
| **library2** | 6 | 5.2 | 147 | 28 | 215.5 | 5,400 | 329 |
| **library3** | 6 | 5.3 | 167 | 16 | 195.5 | 3,800 | 236 |
| **library4** | 6 | 5.5 | 182 | 11 | 192.5 | 3,600 | 234 |
| **library5[1]** | | | | 15 | 106.5 | 1,800 | 347 |
| **library6** | 0 | 4.93 | 222 | 21 | 80 | 503 | 190 |

1. Synopsys PAC utility is unable successfully execute the library5.

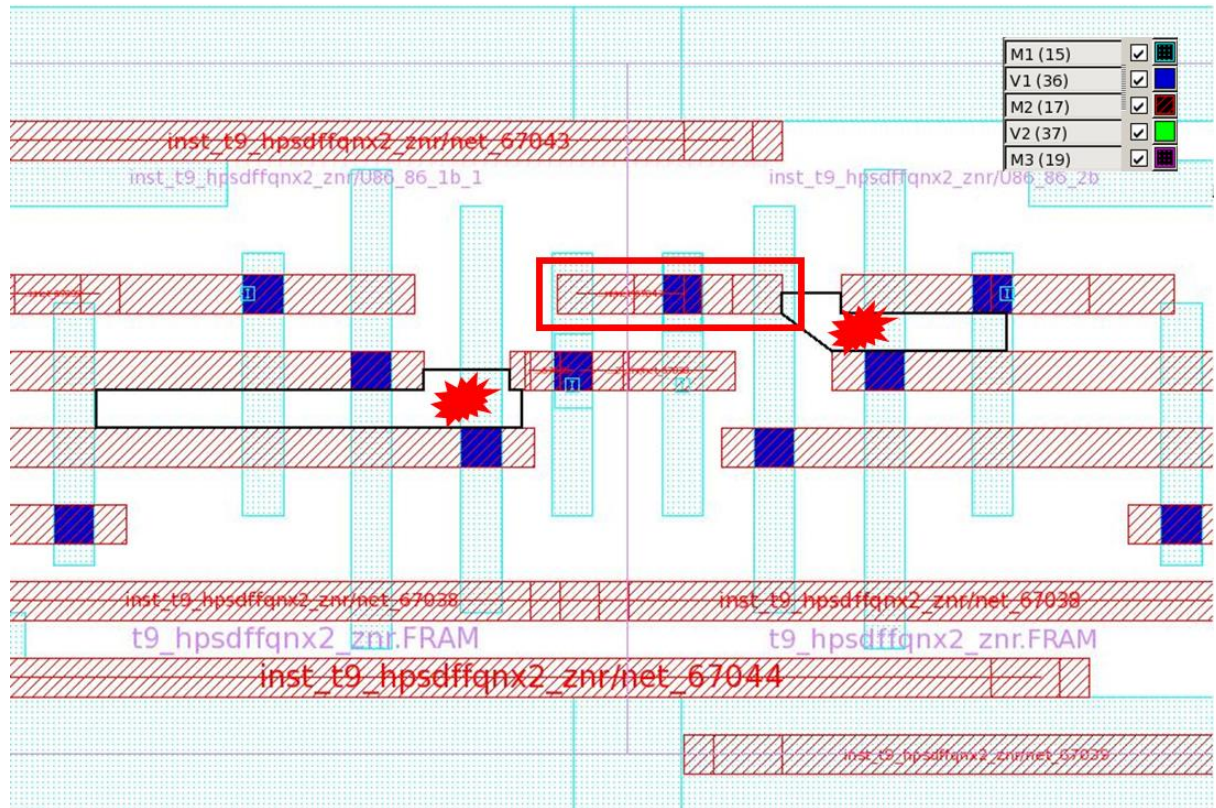

Figure 16. An example illustrating the odd pattern cycle errors near the pin locations due to the insertion of the metal 2 (M2) and via (V1) layers for routing.

*Constraining the Synopsys Pin Access Checking Utility for Improved Standard Cells Library Verification Flow*

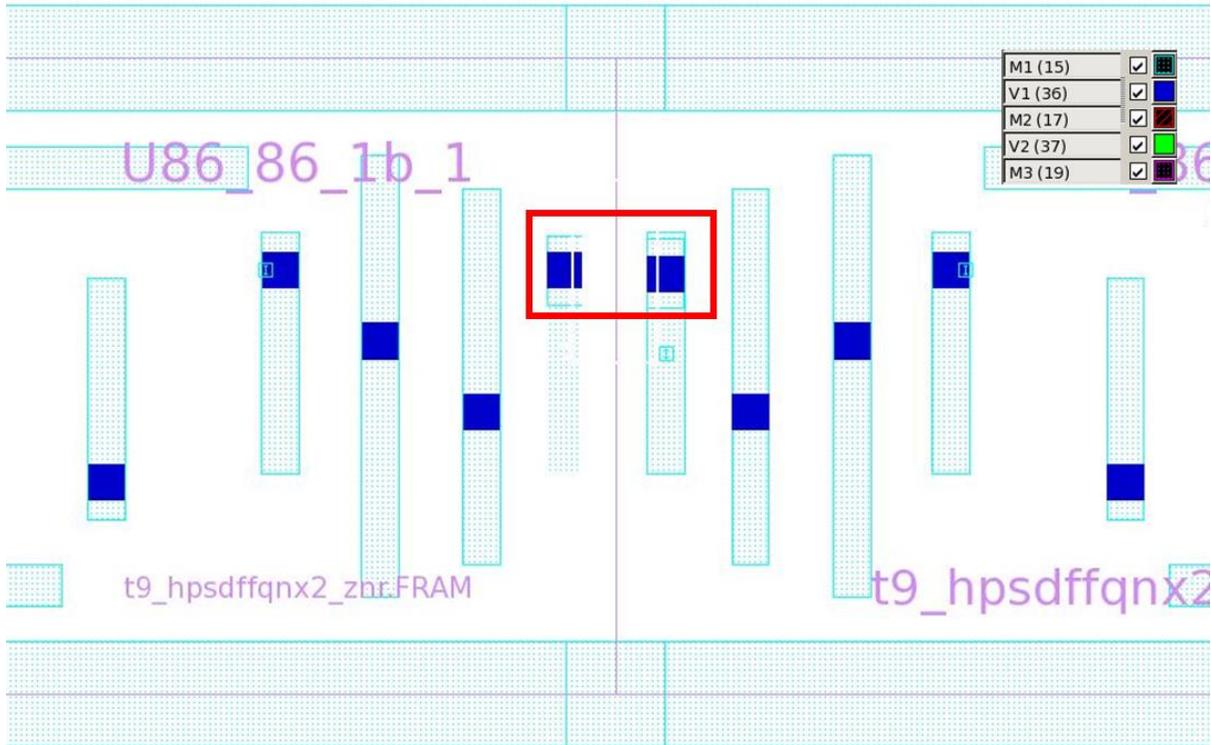

Figure 17. An example illustrating via spacing errors due to the insertion of V1 via layers at the pin locations.

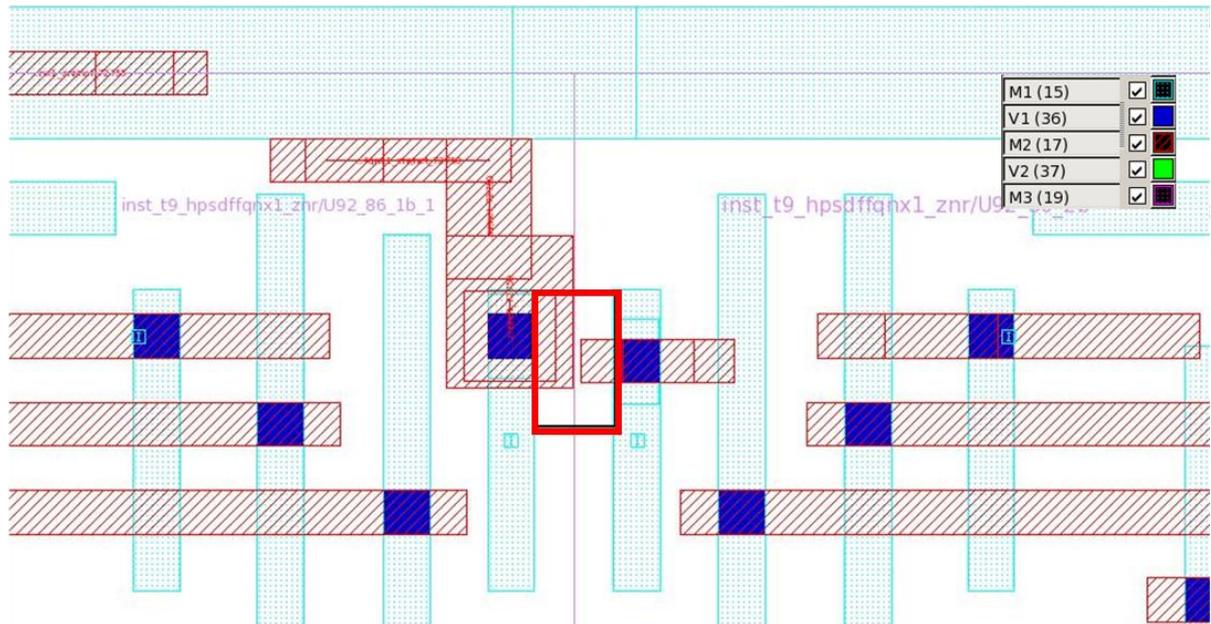

Figure 18. An example illustrating metal 2 (M2) spacing errors due to the insertion of the metal 2 (M2) and via (V1) layers for routing.

## VI.    Future Work

Beside the proposed improvements discussed in the Section IV, we have identified opportunities to enhance the Synopsys PAC methodology in our future work. One example is to modify the milkyway technology file. When we increase 'minEnclosedWidth' to the recommended design margin for better design robustness and manufacturability, we can identify and correct problematic standard cells based the pin locations with the spacing DRC violations. Similarly, when we increase the 'minSpacing' and 'minWidth' to the recommended design margin, it creates routing congestion and increases the probability of detecting pin accessibility issues.

## VII.    Conclusions

In this work, we demonstrated several enhancements to the Synopsys PAC methodology and improved the probability of detecting pin accessibility issues. We recommended using our 'testcell' with the reduced number of cells required, randomizing the pins assignment and router constraints. The significance of the presented methods help library physical layout designer to perform extensive physical verification and identify opportunity to correct the layout before silicon validation phase. As a result, we are expecting a better quality of result for the standard cell design, enabling our customers to achieve best-in-class PPA for their design.